\documentclass[final,3p,times,twocolumn]{elsarticle}

\usepackage[utf8]{inputenc}
\DeclareUnicodeCharacter{03B1}{$\alpha$}
\DeclareUnicodeCharacter{03B3}{$\gamma$}
\DeclareUnicodeCharacter{03B2}{$\beta$}
\usepackage{amssymb}
\usepackage{verbatim}
\usepackage{amsmath}
\usepackage{booktabs}
\usepackage{url}        
\usepackage{hyperref}   
\usepackage{orcidlink}
\makeatletter
\def\ps@pprintTitle{%
     \let\@oddhead\@empty
     \let\@evenhead\@empty
     \let\@oddfoot\@empty
     \let\@evenfoot\@empty
}
\makeatother

\begin{document}

\begin{frontmatter}



\title{Object Detection with Deep Learning for Rare Event Search in the GADGET II TPC}


\author[inst1,inst2,inst3]{Tyler Wheeler}
\author[inst3,inst4]{S. Ravishankar}
\author[inst2,inst1]{C. Wrede}
\author[inst2,inst1]{A. Andalib}
\author[inst2,inst1,inst21]{A. Anthony}
\author[inst1,inst5]{Y. Ayyad}
\author[inst2,inst1]{B. Jain}
\author[inst2,inst1]{A. Jaros}
\author[inst1]{R. Mahajan}
\author[inst2,inst1]{L. Schaedig}
\author[inst2,inst1]{A. Adams}
\author[inst6]{S. Ahn}
\author[inst13]{J.M. Allmond}
\author[inst7]{D. Bardayan}
\author[inst1]{D. Bazin}
\author[inst2,inst1]{K. Bosmpotinis}
\author[inst8]{T. Budner}
\author[inst7]{S.R. Carmichael}
\author[inst6]{S.M. Cha}
\author[inst9]{A. Chen}
\author[inst13]{K.A. Chipps}
\author[inst10]{J.M. Christie\textsuperscript{\orcidlink{0000-0002-8653-8236}}}
\author[inst10]{I. Cox}
\author[inst2,inst1]{J. Dopfer}
\author[inst11]{M. Friedman}
\author[inst12]{J. Garcia-Duarte}
\author[inst1]{E. Good}
\author[inst13,inst10]{T.J. Gray}
\author[inst11]{A. Green}
\author[inst10]{R. Grzywacz}
\author[inst6]{K. Hahn}
\author[inst2,inst1]{R. Jain}
\author[inst14]{E. Jensen}
\author[inst13]{T. King}
\author[inst1]{S. Liddick}
\author[inst12]{B. Longfellow}
\author[inst1]{R. Lubna}
\author[inst16,inst1]{C. Marshall}
\author[inst12]{Y. Mishnayot}
\author[inst17]{A.J. Mitchell}
\author[inst2,inst1]{F. Montes}
\author[inst22,inst1]{T.H. Ogunbeku}
\author[inst2,inst1]{J. Owens-Fryar}
\author[inst13,inst10]{S.D. Pain}
\author[inst1]{J. Pereira}
\author[inst18]{E. Pollacco}
\author[inst19]{A.M. Rogers}
\author[inst2,inst1]{M.Z. Serikow}
\author[inst1]{K. Setoodehnia}
\author[inst1]{L.J. Sun}
\author[inst20]{J. Surbrook}
\author[inst2,inst1]{A. Tsantiri}
\author[inst2,inst1]{L.E. Weghorn}

\affiliation[inst1]{organization={Facility for Rare Isotope Beams, Michigan State University},
            city={East Lansing},
            postcode={48824}, 
            state={Michigan},
            country={USA}}

\affiliation[inst2]{organization={Department of Physics and Astronomy, Michigan State University},
            city={East Lansing},
            postcode={48824}, 
            state={Michigan},
            country={USA}}

\affiliation[inst3]{organization={Department of Computational Math, Science, and Engineering at Michigan State University},
            city={East Lansing},
            postcode={48824}, 
            state={Michigan},
            country={USA}}

\affiliation[inst4]{organization={Department of Biomedical Engineering at Michigan State University},
            city={East Lansing},
            postcode={48824}, 
            state={Michigan},
            country={USA}}

\affiliation[inst5]{organization={Department of Particle Physics at Universidad de Santiago de Compostela},
            city={Santiago de Compostella},
            postcode={E-15706},
            country={Spain}}

\affiliation[inst6]{organization={Center for Exotic Nuclear Studies, Institute for Basic Science},
            city={Daejeon},
            postcode={34126}, 
            country={Republic of Korea}}

\affiliation[inst7]{organization={Department of Physics and Astronomy, University of Notre Dame},
            city={Notre Dame},
            postcode={46556}, 
            state={Indiana},
            country={USA}}

\affiliation[inst8]{organization={Argonne National Laboratory},
            addressline={9700 S. Cass Avenue}, 
            city={Lemont},
            postcode={60439}, 
            state={Illinois},
            country={USA}}

\affiliation[inst9]{organization={Department of Physics and Astronomy, McMaster University},
            addressline={280 Main St W}, 
            city={Hamilton},
            postcode={L8S 4L8}, 
            state={Ontario},
            country={Canada}}

\affiliation[inst10]{organization={Department of Physics and Astronomy, The University of Tennessee, Knoxville},
            city={Knoxville},
            postcode={37996}, 
            state={Tennessee},
            country={USA}}

\affiliation[inst11]{organization={The Racah Institute of Physics, Hebrew University of Jerusalem},
            city={Jerusalem},
            postcode={91904}, 
            country={Israel}}

\affiliation[inst12]{organization={Lawrence Livermore National Laboratory},
            city={Livermore},
            postcode={94550}, 
            state={California},
            country={USA}}

\affiliation[inst13]{organization={Physics Division, Oak Ridge National Laboratory},
            city={Oak Ridge},
            postcode={37831}, 
            state={Tennessee},
            country={USA}}

\affiliation[inst14]{organization={Department of Physics and Astronomy, Aarhus University},
            city={Aarhus},
            postcode={8000}, 
            country={Denmark}}

\affiliation[inst16]{organization={Department of Physics and Astronomy, Ohio University},
            city={Athens},
            postcode={45701}, 
            state={Ohio},
            country={USA}}

\affiliation[inst17]{organization={Department of Nuclear Physics and Accelerator Applications, The Australian National University},
            city={Canberra},
            postcode={ACT 2601},
            country={Australia}}

\affiliation[inst18]{organization={IRFU / DEDIP, CEA Saclay},
            addressline={F91191}, 
            city={Gif-sur-Yvette},
            country={France}}

\affiliation[inst19]{organization={Department of Physics and Applied Physics, University of Massachusetts Lowell},
            city={Lowell},
            postcode={01854}, 
            state={Massachusetts},
            country={USA}}

\affiliation[inst20]{organization={Los Alamos National Laboratory},
            city={Los Alamos},
            postcode={87545}, 
            state={New Mexico},
            country={USA}}

\affiliation[inst21]{organization={Department of Physics and Astronomy, High Point University},
            city={High Point},
            postcode={27268}, 
            state={North Carolina},
            country={USA}}

\affiliation[inst22]{organization={Department of Physics and Astronomy, Mississippi State University},
            city={Mississippi State},
            postcode={39762}, 
            state={Mississippi},
            country={USA}}

\begin{abstract}
In the pursuit of identifying rare two-particle events within the GADGET II Time Projection Chamber (TPC), this paper presents a comprehensive approach for leveraging Convolutional Neural Networks (CNNs) and various data processing methods. To address the inherent complexities of 3D TPC track reconstructions, the data is expressed in 2D projections and 1D quantities. This approach capitalizes on the diverse data modalities of the TPC, allowing for the efficient representation of the distinct features of the 3D events, with no loss in topology uniqueness. Additionally, it leverages the computational efficiency of 2D CNNs and benefits from the extensive availability of pre-trained models. Given the scarcity of real training data for the rare events of interest, simulated events are used to train the models to detect real events. To account for potential distribution shifts when predominantly depending on simulations, significant perturbations are embedded within the simulations. This produces a broad parameter space that works to account for potential physics parameter and detector response variations and uncertainties. These parameter-varied simulations are used to train sensitive 2D CNN object detectors. When combined with 1D histogram peak detection algorithms, this multi-modal detection framework is highly adept at identifying rare, two-particle events in data taken during experiment 21072 at the Facility for Rare Isotope Beams (FRIB), demonstrating a 100\% recall for events of interest. We present the methods and outcomes of our investigation and discuss the potential future applications of these techniques.
\end{abstract}

\end{frontmatter}


\section{Introduction}

\subsection{Nuclear Astrophysics Motivation}\label{phys_motivation}
The light curve of Type I X-ray bursts from neutron stars can provide crucial insights into their mass, radius, and crust elemental abundance. To extract this information, astrophysical simulations of the light curves are compared with observational data acquired using space-based telescopes. However, uncertainties in nuclear reaction rates can significantly impact the accuracy of these simulations. Sensitivity studies have shown that at breakout temperatures around 0.5 GK, the $^{15}$O($\alpha$, $\gamma$)$^{19}$Ne reaction emerges as one of the key reaction rate uncertainties that must be reduced to improve light curve models. Additionally, this reaction ranks among the top reactions influencing the burst ashes in multi-zone models \cite{Cyburt:2016, Fisker_2006}.

Experiment 21072, conducted at the Facility for Rare Isotope Beams (FRIB) at Michigan State University, aims to experimentally constrain this astrophysically relevant \textsuperscript{15}O($\alpha$, $\gamma$)\textsuperscript{19}Ne reaction rate. The results from this experiment will provide cutting edge astrophysical computer simulations with the needed inputs for the accurate modeling of X-ray burst light curves, which will allow for the astrophysics of X-ray bursts to be probed with reduced nuclear physics uncertainties, and potentially the inference of neutron star properties. 

\begin{figure*}
  \centering
  \includegraphics[width=0.9\textwidth]{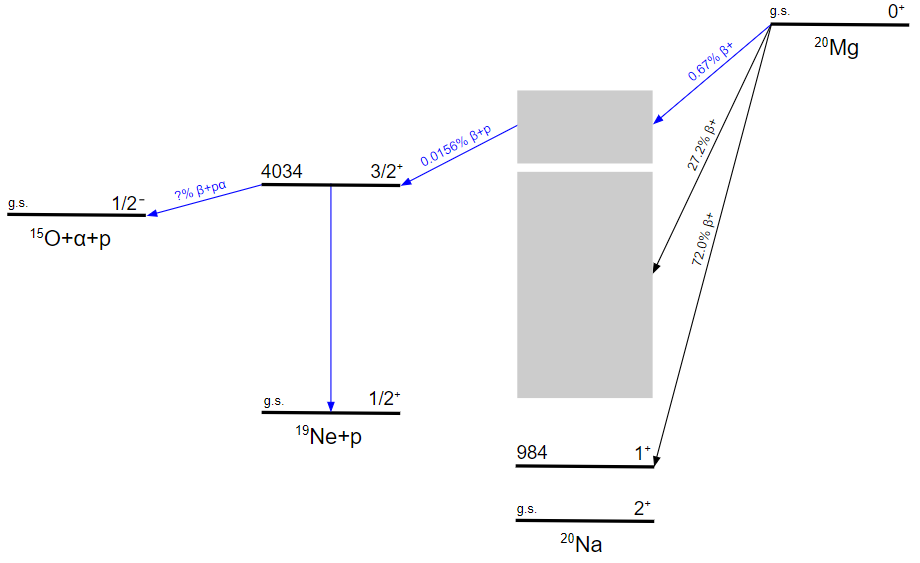}
  \caption{Simplified decay sequence of $^{20}$Mg showing relevant transitions. The $^{15}$O($\alpha$, $\gamma$)$^{19}$Ne reaction rate is primarily influenced by a resonance with a well-established center-of-mass energy of 506 keV, corresponding to the 4034 keV state in $^{19}$Ne \cite{tyler_thesis}.}
  \label{fig:20MgDecay}
\end{figure*}

This reaction can be measured indirectly via the decay sequence of \textsuperscript{20}Mg (Fig. \ref{fig:20MgDecay}). A single \textsuperscript{20}Mg($\beta p \alpha$)\textsuperscript{15}O event through the key \textsuperscript{15}O($\alpha$, $\gamma$)\textsuperscript{19}Ne resonance yields a characteristic signature: the near simultaneous emission of a proton and alpha particle (center-of-mass energies: $E_p$ = \(1.21^{+0.25}_{-0.22}\) MeV, $E_\alpha$ = 506.2(17) keV) \cite{PhysRevC.96.032801, PhysRevC.99.065801}. To resolve the proton-alpha coincidence events the Proton Detector from the GAseous Detector with GErmanium Tagging (GADGET) detection system has been upgraded into a time projection chamber (TPC). This upgraded system is referred to as the GADGET II TPC \cite{Friedman:2019, PhysRevC.110.035807}. 

The GADGET II TPC (further discussed in Sec. \ref{gadget}) allows for full 3D reconstructions of decay events occurring in the detector's active volume, and proton-alpha coincidence events will have a unique 3D topology. However, challenges arise given that the $\beta p$ feeding intensity for the key resonance is extremely low, measured at 0.0156(38)\% \cite{PhysRevC.96.032801} or 0.0149(35)\% \cite{PhysRevC.99.065801}. Additionally, strong upper limits for the alpha branching ratio have been found to be $6 \times 10^{-4}$ \cite{Rehm2003}, and $4.3 \times 10^{-4}$ \cite{PhysRevC.67.065808}, with $\sim3\times 10^{-4}$ being a reasonable estimate for the expected value \cite{Tan2007}. As a result, $^{20}$Mg($\beta p \alpha$) events are exceptionally rare, making the task of locating one within the vast amount of TPC data highly challenging.

In most cases, recorded events from the $^{20}$Mg decay sequence will involve protons from $^{20}$Mg($\beta p$) (29.9(11)\% of decays) or alphas from the $^{20}$Na($\beta \alpha$) daughter decay (19.9(2)\% of decays). However, a significant source of background contamination arises from the decay of $^{21}$Mg($\beta p \alpha$) (center-of-mass energies: $E_p = 919(18)$ keV, $E_\alpha = 882(18)$ keV \cite{LUND2015356}), which, although rare with a branching ratio per $^{21}$Mg decay of 0.016(3)\%, is still orders of magnitude more frequent than the desired $^{20}$Mg($\beta p \alpha$) events. Distinguishing between these two decay sequences requires statistically quantifiable determination of the energy sharing between the particles. While we are exploring the use of the Markov Chain Monte Carlo (MCMC) method to analyze 3D tracks and extract particle energies from two-particle decays, this work is ongoing. The main focus of this paper is the development of a detection framework capable of identifying rare, two-particle decays within the vast TPC dataset, enabling further analysis to confirm the isotopic origin of the two-particle events.

\subsection{\textit{Deep Learning Motivation}}

Techniques involving the search for rare events have been explored extensively in particle physics \cite{PhysRevLett.121.241803, Crispim_Rom_o_2021, PhysRevD.105.095004}, however, these approaches are often used for general anomaly detection, where any deviation from background can be flagged. In contrast, the challenge at hand involves using a TPC to identify a very specific rare event—the proton-alpha coincidence. With the goal of leveraging machine learning to sift through this extensive dataset ($\sim$ 68 million recorded events), we proposed feeding 2D projections of each 3D track into a 2D convolutional neural network (CNN), a strategy that has been used successfully in both nuclear and particle physics TPC applications \cite{KUCHERA2019156, Acciarri_2017}. While this problem may seem ideally suited for a 3D CNN, dense 3D networks are far too computationally expensive for large TPC datasets. Additionally, the availability of high-quality pre-trained 2D CNN models makes them a more attractive choice. However, there are several significant challenges that need to be addressed in this case, especially due to the importance of accurately identifying all relevant events in a rare event search.

One issue is that 2D projections may fail to resolve both particles if the event track is nearly perpendicular to the TPC detection plane. However, 2D models can still be used by exploiting the different data modalities captured by the TPC. All of the relevant data from a 3D track can be effectively represented through a combination of 2D, and 1D data quantities (see Sec. \ref{projections}). This dimensionality reduction allows for the efficient use of 2D CNNs with no loss in topology uniqueness. 

The rare nature of proton-alpha events introduces an additional challenge: an absence of real data for training a network. This issue is addressed by relying solely on simulations for training (Sec. \ref{simsec}), but models often perform poorly on real datasets when trained only on simulations \cite{KUCHERA2019156}. To combat this, a two-step approach is employed. First, the simulations incorporate robustification techniques to broaden the parameter space and improve the generalization of the model (Sec. \ref{robust}). Second, instead of employing a standard CNN classifier, the framework utilizes 2D object detectors (Sec. \ref{object_det}), which focus on identifying key features of two-particle events, specifically, the presence of a double Bragg peak in the projected track. When combined with 1D histogram peak detection on the time projection for highly perpendicular events, this method effectively captures two-particle events in real datasets when trained only on simulated data.

The full detection framework described in this paper functions as a data processing pipeline,  designed to identify proton-alpha coincidence events efficiently. The framework significantly improves detection reliability while greatly reducing the manual effort required for finding these events. The following sections detail all the components of the framework and present the results from its deployment on data from experiment 21072. 

\subsection{\textit{The GADGET II TPC}}\label{gadget}
\begin{figure}
\includegraphics[width=\linewidth]{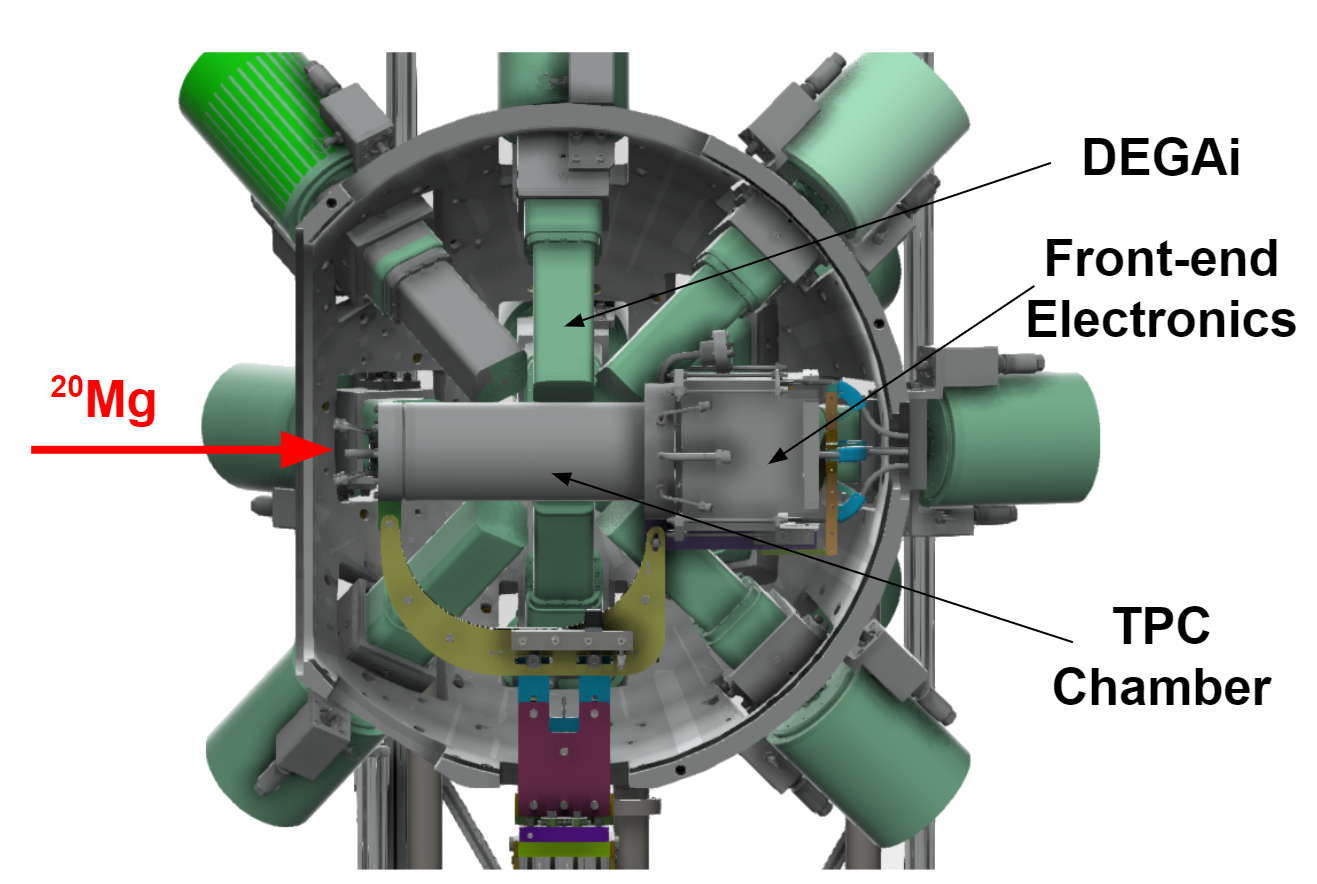}
\caption{\label{20Mg Decay} CAD drawing of the GADGET II detection system showing the direction a radioactive beam enters the TPC chamber. The GADGET II TPC is seen inside of the DEGAi germanium array \cite{tyler_thesis}.}
\label{GADGET_Sys}
\end{figure}

\begin{figure}[htbp]
\centering
\includegraphics[width=0.65\linewidth]{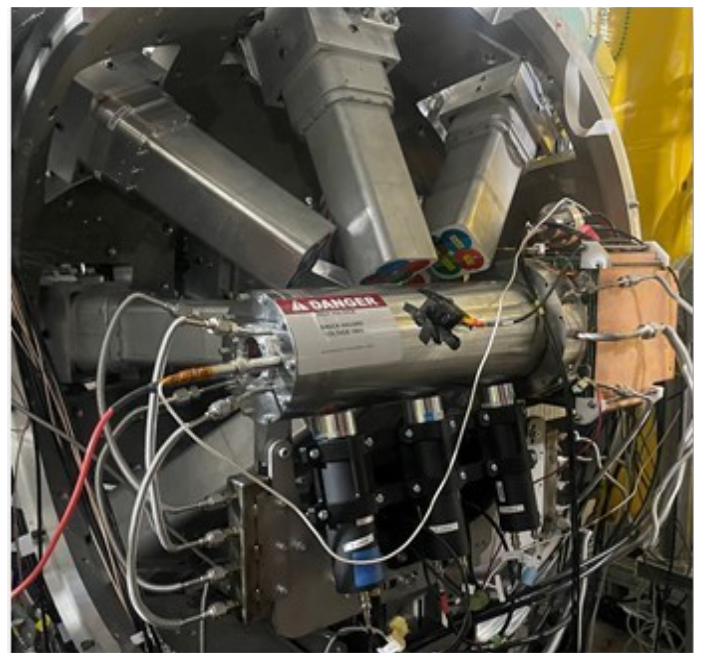}
\caption{\label{20Mg Decay} Full GADGET II experimental setup used during FRIB experiment 21072 \cite{tyler_thesis}.}
\label{GADGET_Sys_exp}
\end{figure}

In its FRIB beam-line configuration, GADGET II has two key elements: The DEcay Germanium Array initiator (DEGAi) \cite{FRIB2023} for high efficiency $\gamma$ detection, and the TPC (see Figs. \ref{GADGET_Sys}, and \ref{GADGET_Sys_exp}). The DEGAi data is not relevant for event classification in this case, so the focus of this section will be on the TPC and other ancillary components.

The TPC is filled with P10 gas (90\% Argon, 10\% Methane) which serves a dual purpose: acting as a beam stop for fast rare isotope beams and as a detection medium through ionization electrons generated by the kinetic energy of particles released by subsequent decays. When exposed to a uniform electric field, these ionization electrons drift towards the detection pads. The electric field is divided into two distinct regions: the drift region and the proportional amplification region. In the drift region, the electric field is set to $\sim$125 V/cm to ensure ionized electrons traverse the active volume without liberating any additional charge. Conversely, the proportional amplification region features a very strong electric field, $\sim$35 kV/cm over a 128 $\mu$m gap, triggering a Townsend avalanche, amplifying the event signals in the process. Following this the charge cluster is collected by a resistive anode, which then induces a charge on the Micromegas that is read-out by the front-end electronics \cite{POLLACCO201881}.

\begin{figure}
\includegraphics[width=\linewidth]{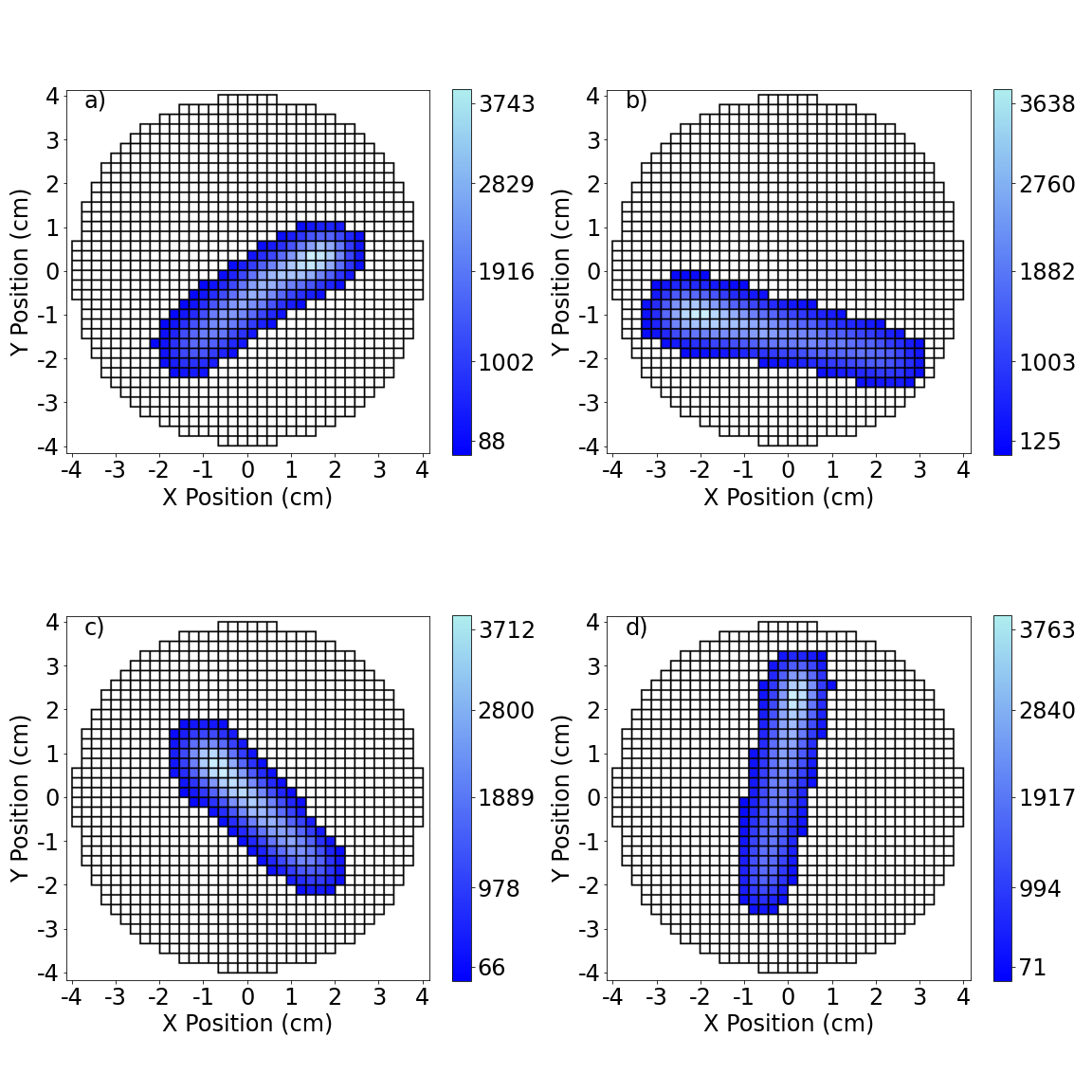}
\caption{\label{PadPlane_track} 2D MicroMegas pad plane images of \textsuperscript{220}Rn alpha tracks (6.288 MeV) inside the GADGET II TPC \cite{tyler_thesis}.}
\label{2D_tracks}
\end{figure}

\begin{figure*}
\centering
\includegraphics[width=0.9\linewidth]{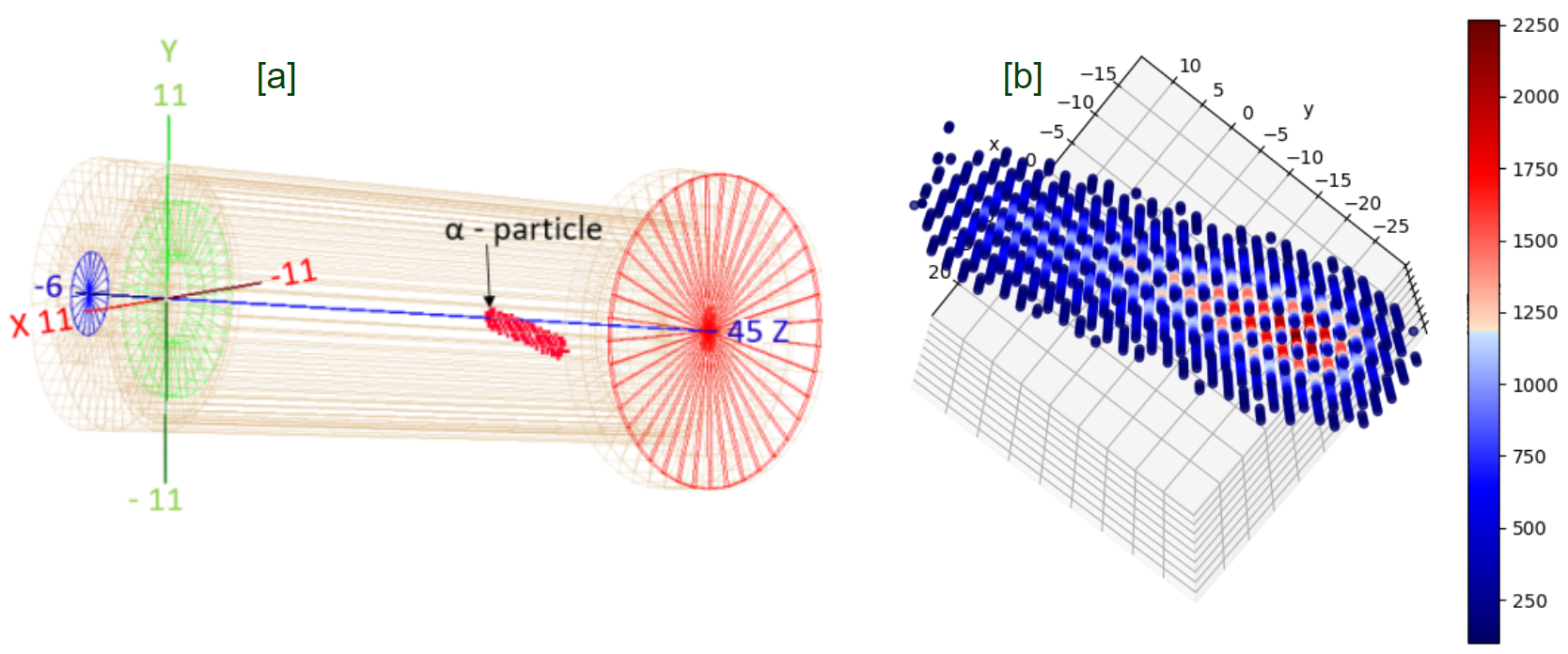}
\caption{\label{TPC_track3D} a) 3D reconstruction of a real 6.288 MeV \textsuperscript{220}Rn alpha track inside the GADGET II TPC. b) Closeup of the reconstruction with energy deposition showing a characteristic Bragg peak \cite{tyler_thesis}.}
\end{figure*}

Equipped with over a thousand 2.2 x 2.2 mm\textsuperscript{2} pads, the resistive Micromegas offers the spatial resolution necessary to generate a 2D image of decay events. An example of these 2D track projections can be seen in Fig. \ref{2D_tracks}. The third dimension is derived from the drift time distribution of the incoming electrons, enabling a comprehensive 3D reconstruction of decay events within the TPC's active volume. A full 3D reconstruction of a decay event is shown in Fig. \ref{TPC_track3D}. For a more in-depth exploration of GADGET II and its functionalities, refer to \cite{PhysRevC.110.035807}.

\section{Simulating Training Data}\label{simsec}
The absence of hand-labeled events of interest in the real dataset presents a challenge. To address this issue training data is created using high-quality simulations of events of interest. These simulations are carefully tailored to represent the relevant two-particle decay events in the GADGET II TPC. While it is often true that models trained only on simulated data will perform poorly on real data \cite{KUCHERA2019156}, this problem is overcome through robustification and a focus on feature detection (see Secs. \ref{robust} and \ref{obdet_bragg}).

\subsection{\textit{ATTPCROOTv2}}\label{attpcroot}
\begin{figure*}
\centering
\includegraphics[width=0.9\linewidth]{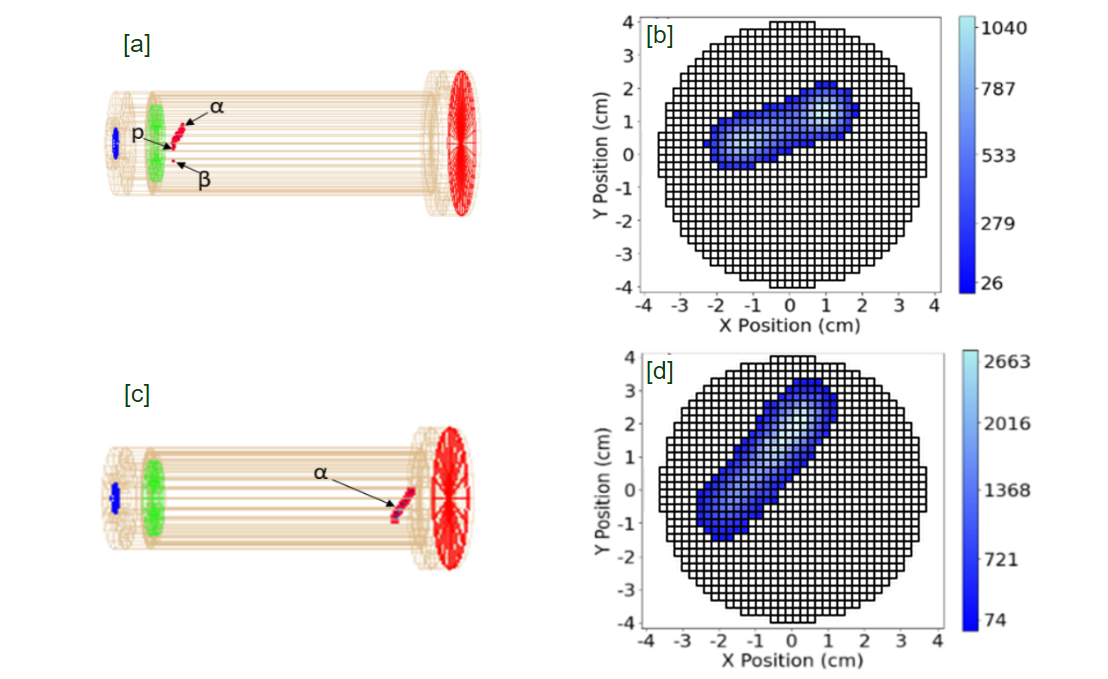}
\caption{\label{alphaproton} Panels (a) and (b) depict an ATTPCROOTv2 simulation for the $^{20}$Mg($\beta$p$\alpha$)$^{15}$O decay sequence in the GADGET II TPC, showcasing both a 3D render and 2D projection. Panels (c) and (d) present the ATTPCROOTv2 simulation for the $^{220}$Rn $\alpha$-decay sequence in similar formats.}
\end{figure*}

Simulations are generated using the ATTPCROOTv2 \cite{Ayyad2023} data analysis framework, which is built upon the FairSoft and FairRoot packages. This comprehensive framework leverages established nuclear physics libraries and a suite of physics generators to accurately simulate decay events. The ATTPCROOTv2 framework offers users the flexibility to unpack, analyze, and even design a custom geometry for conducting event-by-event simulations through a virtual Monte Carlo package. Each simulated event can represent specific decay sequences, such as $^{20}$Mg($\beta$p$\alpha$)$^{15}$O and $^{220}$Rn $\alpha$-decay, as illustrated in Fig. \ref{alphaproton}. A significant advantage of the ATTPCROOTv2 framework is its capability to produce simulation outputs that mirror real data, ensuring consistency in subsequent analyses.

Following the generation of simulated data, pulse shape analysis processes the simulated readouts from the Micromegas pads. Advanced pattern recognition algorithms evaluate each event, enabling the tracking of particle trajectories within the detector. The Monte Carlo simulations are facilitated using the Geant4 toolkit, while the HDF5 library ensures appropriate data formatting \cite{AYYAD2020161341}.

\subsection{\textit{Issues with 2D Projections}}\label{projections}

In the context of our study, the events of interest are proton-alpha coincidence events with a combined energy of approximately 1.7 MeV (1.2 MeV from the proton and 0.5 MeV from the alpha). While the 3D topology of this type of event is distinct, its 2D projection only retains uniqueness if the track does not have a large component perpendicular to the projection plane.

\begin{figure}
\centering
\includegraphics[width=0.8\linewidth]{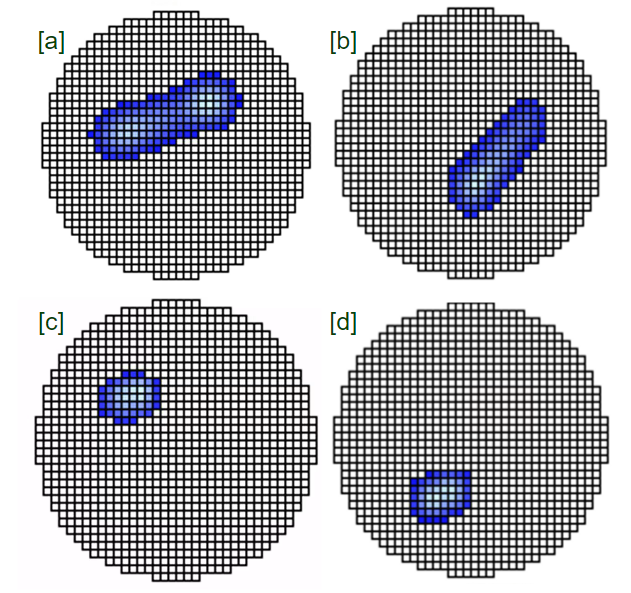}
\caption{\label{angle} Comparison of simulated 2D projections of particle events on the pad plane. (a) Proton-alpha coincidence event parallel to the pad plane, showcasing distinct energy depositions at both ends. (b) Isolated proton event parallel to the pad plane. (c) Proton-alpha coincidence event perpendicular to the pad plane, appearing as an indistinct disc of energy deposition. (d) Isolated proton event perpendicular to the pad plane, similarly manifesting as an indistinct disc of ionization.}
\end{figure}

Consider, for example, two potentially confusable classes: a proton-alpha coincidence event and an isolated proton event. When these events occur parallel to the pad plane, the proton-alpha coincidence event can be readily identified by the pronounced energy deposition at both ends of the track, as illustrated in Fig. \ref{angle}a-b. One peak arises from the proton's characteristic Bragg peak, while the other results from the point-like energy deposition of the low-energy alpha. Conversely, if these events were to decay perpendicular to the pad plane, they would both appear as circular charge distributions, rendering them virtually indistinguishable (refer to Fig. \ref{angle}c-d).

\begin{figure*}
\centering
\includegraphics[width=0.9\linewidth]{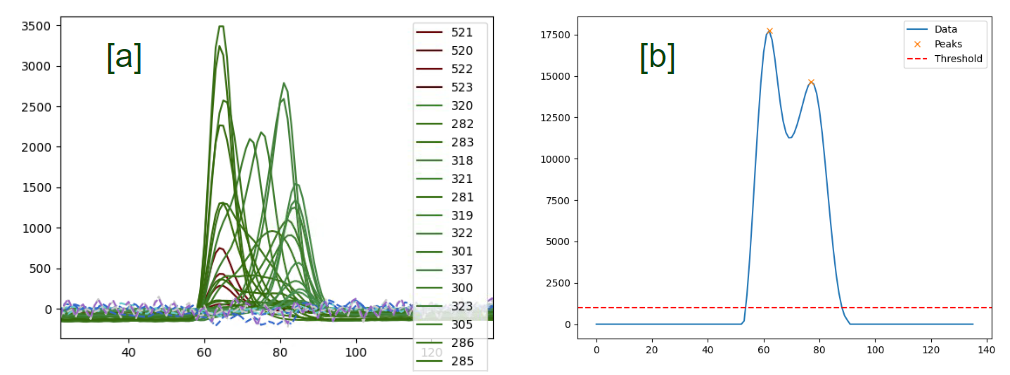}
\caption{\label{traces} (a) Pad traces from a proton-alpha coincidence event. The y-axis is the digitized signal value and the x-axis is the time bucket index. The labels on the right indicate which channel on the pad plane each trace corresponds to. (b) Time projection, generated by summing all pad signals for a proton-alpha event and applying smoothing. A peak finding algorithm flags the two prominent peaks.}
\end{figure*}

\begin{figure}
\centering
\includegraphics[width=0.8\linewidth]{{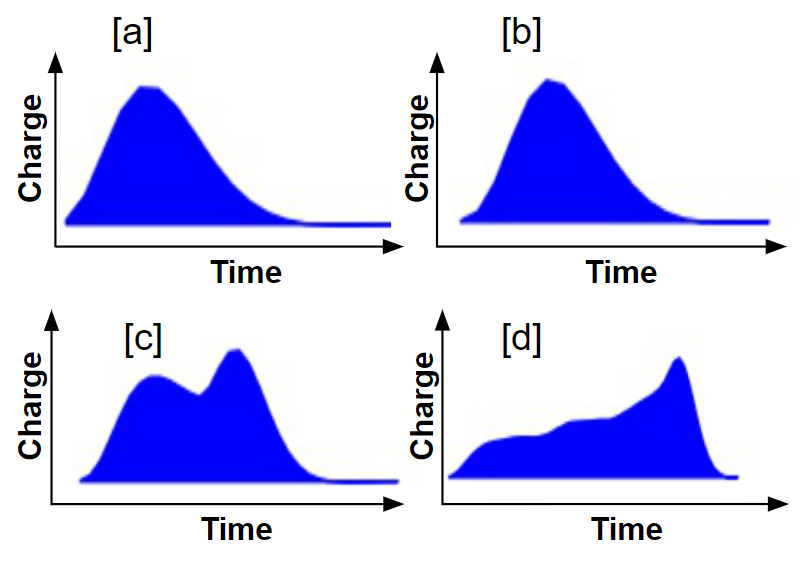}}
\caption{\label{timeprojection} Simulated time projection representations corresponding to the particle events. (a) Time projection of a proton-alpha coincidence event occurring parallel to the pad plane. (b) Time projection of an isolated proton event parallel to the pad plane. (c) Distinct double-peaked signature of the proton-alpha coincidence event occurring perpendicular to the pad plane. (d) Time projection of an isolated proton event perpendicular to the pad plane \cite{tyler_thesis}.}
\end{figure}

This orientation-dependent ambiguity implies that as 2D projected events become increasingly perpendicular relative to the pad plane, they become more susceptible to misclassification. However, an important contrast emerges when considering the time projection. The time projection represents the integrated charge of the track over time, and it is generated by summing all of the trace signals from every pad for an event (see Fig. \ref{traces}). In this time projection representation, parallel events appear indistinguishable, but perpendicular alpha-proton tracks exhibit a double-peaked signature, as depicted in Fig. \ref{timeprojection}c. This observation underscores the complementary nature of the 2D and 1D datasets.

A deeper discussion of the preprocessing involved for the generation of these 2D and 1D datasets can be found in Sec. \ref{preprocessing}.

\subsection{\textit{Robustification through Parameter Variation}}\label{robust}
One difficulty with training a machine learning network on simulated images to classify real images is that the simulations can never perfectly capture the real data. This manifests as a distribution shift between the simulated and real data, and works to diminish model performance. However, this issue can be combated through the use of robustification. The notion of robustification comes from work on adversarial perturbations, where changes are made to inputs so as to deliberately fool a machine-learned model. To shield against these attacks, researchers have developed the notion of adversarial robustness \cite{goodfellow2015explaining}. Although the distribution shift from environmental perturbations is not deliberately adversarial, we can learn from the ideas of adversarial robustness to strengthen our network's performance against such shifts. 

Building robustification into a model often involves introducing perturbations into the training set. This increases training time, but makes the model more robust to distribution shifts. In our case, these perturbations relate to imprecisely known physics parameters embedded in our simulations. For instance, the transverse diffusion coefficient used in the simulations may be appreciably different than the true value, which could lead to a different number of pads firing for a given event, and thus, misclassification could result.

To address the challenge posed by potential discrepancies between simulations and real data, a two-fold strategy is adopted. Initially, simulations are tuned to align closely with the real data. This is achieved by isolating specific high-statistics decay events in the real dataset (800 keV $^{20}$Mg($\beta p$) events, for example), and simulating these same events. A specialized program was developed that treats each simulation parameter as a free variable. This program iteratively adjusts these parameters until there is a close match between the simulated and real data, ensuring that simulations are as representative of the real-world scenarios as possible. The parameters include: transverse diffusion, longitudinal diffusion, gas gain, gas pressure, energy, pad threshold, and charge dispersion. 

\begin{figure}
\includegraphics[width=\linewidth]{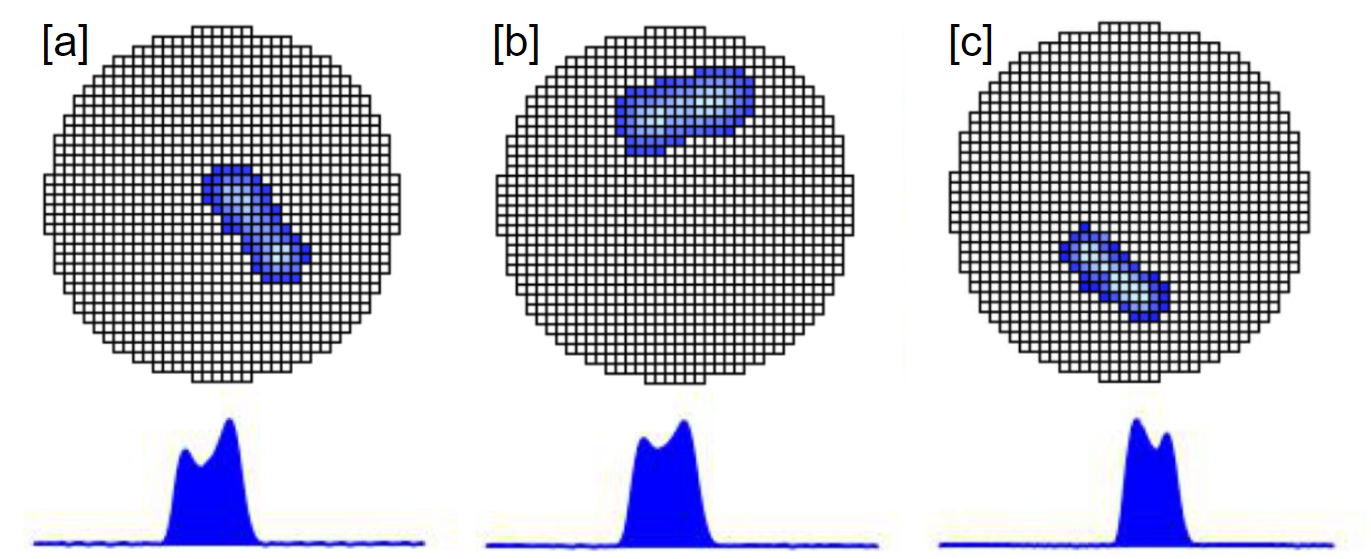}
\caption{\label{varied} Comparison of simulated proton-alpha events with varying physical parameters and their corresponding time projections. (a) Base simulation parameters yielding a standard track representation. (b) Simulation with increased transverse diffusion. (c) Simulation with heightened gain and pad threshold \cite{tyler_thesis}.}
\end{figure}

After determining the optimal parameters, the next phase involves parameter variation. The central values obtained from the tuning process serve as means for Gaussian distributions. During event simulation, parameter values for each event are randomly sampled from these distributions, with the range extending up to 2 standard deviations from the mean. This broadened parameter space paints a more comprehensive and diverse picture of potential event appearances, as shown in Fig. \ref{varied}. As discussed in the following section, by integrating these parameter-varied events into a training dataset, models are exposed to a wider spectrum of data scenarios. This exposure equips the models to discern more generalized and invariant features, thereby enhancing performance.

\section{Object Detection with Deep Learning}\label{object_det}
This section focuses on the 2D TPC image data. However, it is important to note that our rare-event detection framework ultimately combines insights from both 2D track projections and 1D time projections to make final decisions on event classification. This integration of datasets is discussed in detail in Sec. \ref{full_framework}, where the complete pipeline is presented.

Given that TPC particle tracks can be visualized as 2D images, as is illustrated in Fig. \ref{PadPlane_track}, CNNs emerge as among the most suitable machine learning algorithms for our dataset. CNNs are specialized deep learning algorithms optimized for image analysis tasks such as classification, segmentation, and object detection \cite{NIPS2012_c399862d, simonyan2015deep}. These algorithms utilize a hierarchical structure, where each layer captures increasingly abstract features of the input image. The early layers typically identify simple, low-level patterns, while deeper layers combine these features to detect more complex structures. 

\subsection{Object Detection with CNNs}\label{obdet_bragg}
Object detection in computer vision involves two primary challenges: identifying the presence of objects within an image (classification) and accurately delineating the location of these objects by drawing bounding boxes around them (localization) \cite{8627998}.

Methods like the Region-based Convolutional Neural Network (R-CNN) and its variants, Fast R-CNN and Faster R-CNN, have shown that CNNs can achieve remarkable accuracy in object detection due to their capability to perform both classification and regression tasks simultaneously \cite{7485869}. These methods operate by first generating region proposals—potential bounding boxes where objects might be located—and then applying CNNs to these regions to classify the objects and refine their locations. 

Developments in deep learning methods, particularly for tasks such as 2D Bragg peak detection, have demonstrated improvements in both accuracy and computational efficiency compared to traditional techniques like 2D pseudo-Voigt fitting \cite{Liu2022}. Inspired by these advances, we propose utilizing an object detector to localize particle tracks and identify any Bragg peaks within the event track as part of a proton-alpha rare-event search framework.

\subsection{Faster R-CNN for Rare Event Searches}\label{obdet_bragg} 
For our rare-event detection framework, we employ the Faster R-CNN, a deep learning model that integrates region proposal and object detection into a single, end-to-end trainable architecture. The Faster R-CNN is particularly well-suited for this task due to its ability to balance high detection accuracy with computational efficiency, utilizing a Region Proposal Network (RPN) to generate object proposals and a detection head for classification and bounding box refinement.

Detailed descriptions of the Faster R-CNN architecture, including its region proposal, classification, and bounding box regression components, can be found in the literature \cite{8627998, 7485869}. In this work, we adapt the Faster R-CNN to identify and localize particle tracks and Bragg peaks within 2D TPC images, forming a key component of our rare-event detection pipeline.

\subsection{Training an Object Detector}
Achieving optimal model performance with an object detection model, such as the Faster R-CNN, requires careful data preparation, hyperparameter tuning, and validation. This section outlines the process undertaken in our implementation, beginning with the generation and annotation of simulated data, followed by model initialization and training, and culminating in the validation and refinement of the model. 

\begin{itemize}
\item \textbf{Data Preparation:} Simulated 2D track projections with varied parameters are generated using the ATTPCROOTv2 software (Secs. \ref{attpcroot} and \ref{robust}). These simulations are annotated using the VGG Image Annotator (VIA) software \cite{dutta2019via}, developed by the Visual Geometry Group (VGG) (see labeled training examples in Fig. \ref{labels}). The annotations are exported as a CSV file, which includes the bounding box coordinates and class labels. These coordinates and labels are then processed, where the annotations are organized by image and converted into tensors that can be directly fed into the model. To further enhance the robustness of the model, data augmentations such as random horizontal/vertical flipping, and random rotations are applied during the training phase. 

\begin{figure*}
\centering
\includegraphics[width=0.9\textwidth]{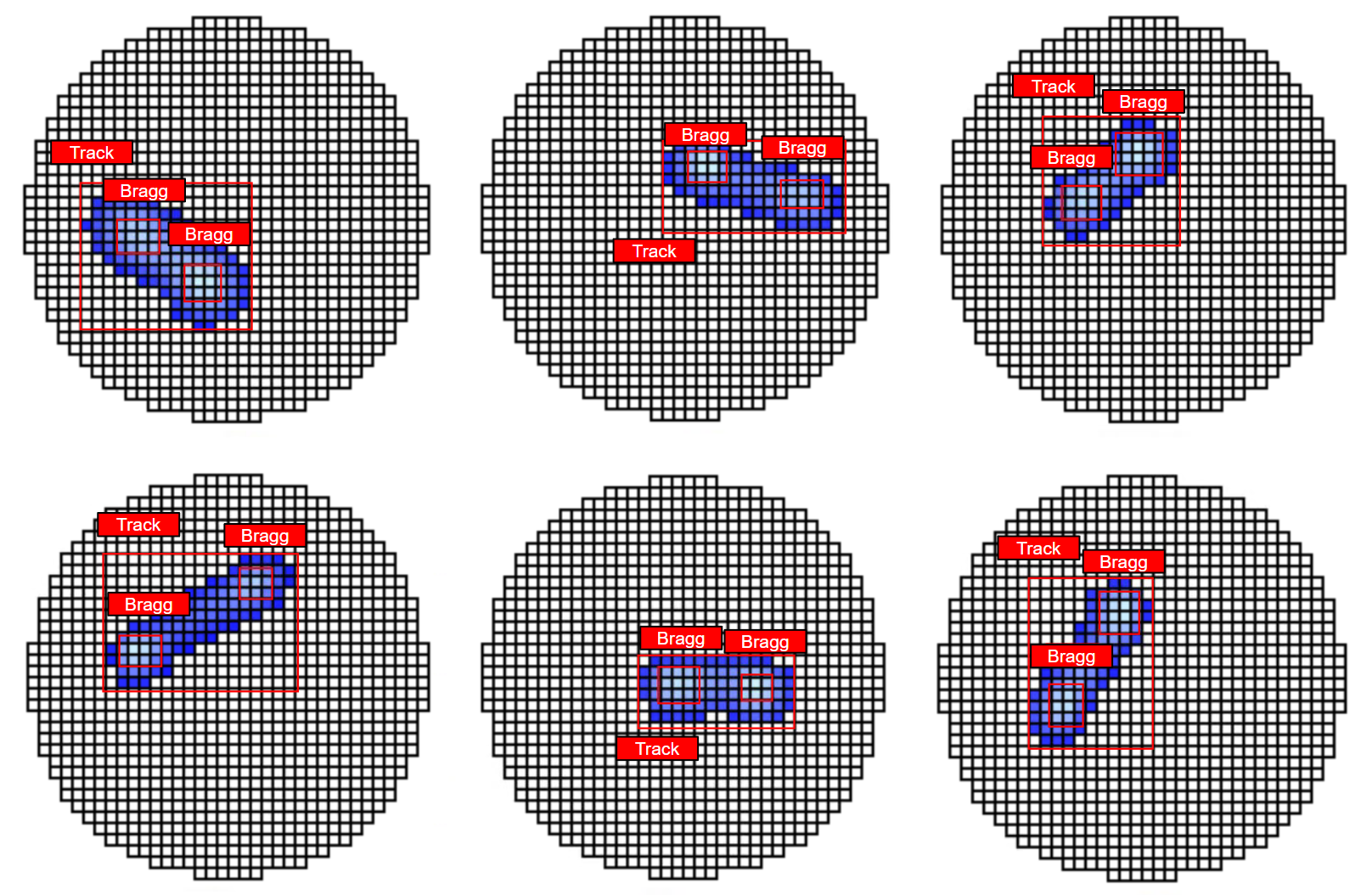}
\caption{\label{labels} Examples of parameter-varied simulations of proton-alphas used for object detector training. Ground-truth bounding box annotations were generated using VIA software.}
\end{figure*}

\item \textbf{Model Initialization:} We utilize a pre-trained Faster R-CNN model with a ResNet-50 backbone, fine-tuning it for our specific task. The model is initialized with weights trained on the COCO dataset \cite{lin2015microsoftcococommonobjects}, and we replace the pre-trained classification head with a new one tailored to our dataset, which has \( 3 \) classes (background, track, and Bragg peak). The fine-tuning allows the model to leverage the rich feature representations learned on a large-scale dataset while adapting to the specific characteristics of our task.

\item \textbf{Training Procedure:}  The Faster R-CNN is fine-tuned using an end-to-end training approach to optimize performance on our dataset. First, the entire model—including the backbone network (e.g., ResNet-50), the RPN, and the detection head—is trained simultaneously. This joint optimization allows all components to adapt cohesively, leveraging the interdependencies between feature extraction, region proposal generation, and object detection. 

To maintain the stability of pre-trained features while allowing significant updates where needed, differential learning rates are employed during training. A lower learning rate is applied to the backbone network to preserve its learned feature representations, while higher learning rates are assigned to the RPN and detection head to facilitate more substantial adaptation to the new dataset. This approach balances the preservation of useful pre-trained features with the flexibility to learn new patterns specific to our data.

\item \textbf{Optimization and Learning Rate Scheduling:} Hyperparameters are selected using the hyperparameter optimization software, Optuna \cite{10.1145/3292500.3330701}. The model is trained using Stochastic Gradient Descent (SGD) with differential learning rates: a learning rate of \(0.001\) is applied to the backbone network (ResNet-50), while a higher learning rate of \(0.005\) is used for the RPN and detection head. 

Both learning rates are scheduled to decrease using a StepLR scheduler, which reduces the learning rates by a factor of \(0.1\) every \(3\) epochs. This scheduling helps the model converge more smoothly by allowing it to take larger steps initially and then fine-tuning as it approaches a local minimum. The momentum is set to \(0.9\), and a weight decay of \(0.0005\) is applied to regularize the model. The batch size is \(2\), and the model is trained to convergence for a total of \(50\) epochs.

\item \textbf{Loss Calculation:} During each training step, the model computes both classification and regression losses for the anchors. The classification loss is responsible for ensuring that the model correctly identifies the presence of objects, while the regression loss ensures that the predicted bounding boxes align closely with the ground truth. The total loss is then backpropagated through the network, and the model's weights are updated accordingly. This multi-task loss function is crucial for the simultaneous optimization of both object detection and localization. The classification loss is a multi-class cross-entropy loss, while the regression loss is a smooth \( L_1 \) loss (Huber loss).

\item \textbf{Validation and Model Saving:} After each epoch, the model is evaluated on a validation set, which comprises \( 20\% \) of the dataset, while the remaining \( 80\% \) is used for training. The average validation loss is tracked throughout the training process. If the validation loss improves (i.e., decreases) compared to the previous best, the weights are saved. 

\end{itemize}
This training regimen ensures that the model is both accurate and efficient, leveraging the power of CNNs for object detection. By using parameter-varied simulations, incorporating data augmentation, fine-tuning, careful hyperparameter selection, and regular validation, we maximize the model's ability to generalize and perform well on new, unseen data.

\section{Full Framework Implementation for Rare Event Detection}\label{full_framework}
The rare event detection framework is a data processing pipeline designed to identify proton-alpha coincidence events. It combines 2D object detection with 1D histogram peak detection to capture these rare events. 

\subsection{Data Selection: Search Region}\label{sec:search_region}
To reduce the amount of data from Experiment 21072 that must be analyzed for rare events, a targeted search region is defined based on the specific particle energies for our events of interest. The proton energy for these events is expected to be \(1.21^{+0.25}_{-0.22}\) MeV \cite{PhysRevC.99.065801}. Protons with this energy will travel $\sim$2-3 cm in 800 Torr P10 gas. The alpha particles in the coincidence event will deposit around 506 keV over a very short range ($<$4 mm) (ranges obtained via SRIM \cite{ZIEGLER1988215}). 

\begin{figure}
\centering
\includegraphics[width=0.5\textwidth]{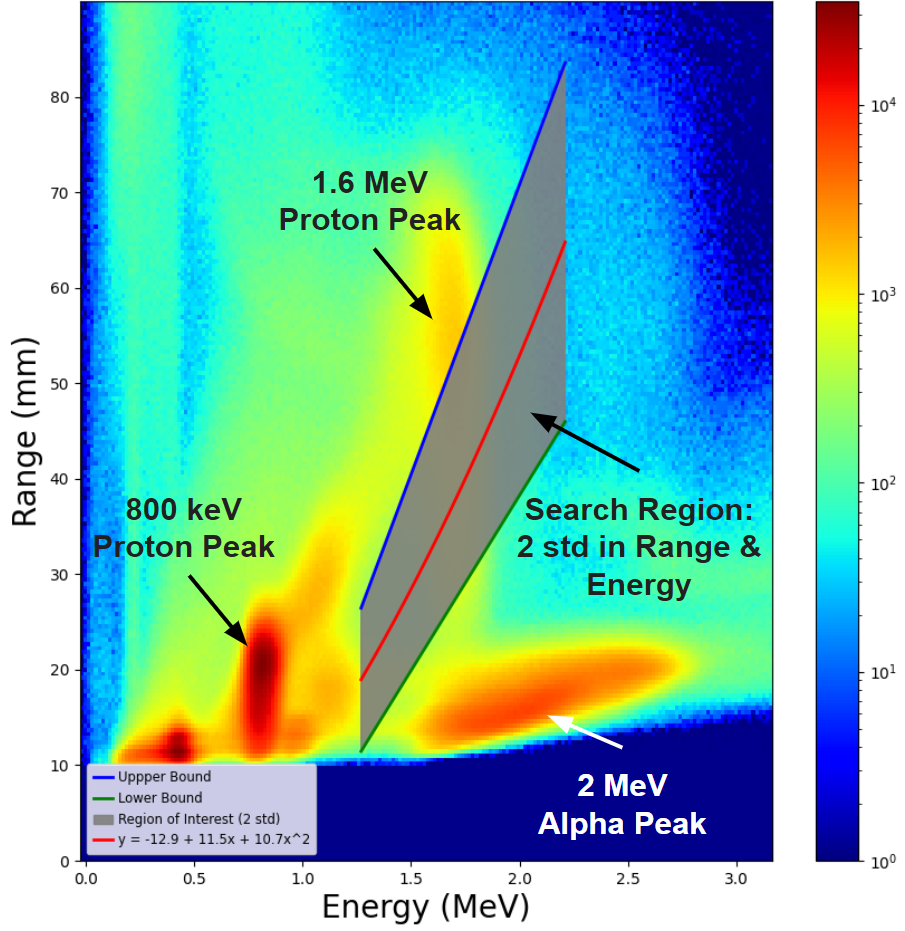}
\caption{Range vs. Energy plot for Experiment 21072 (170 hours of data) shown on a log scale. This plot demonstrates that different decay events will form dense clusters, with proton and alpha decay events forming distinct bands. The plot highlights the defined search region between these bands, covering 2 standard deviations in both range and energy.}
\label{fig:search_region}
\end{figure}

The ratio of track length to total energy deposited is a key characteristic that distinguishes $p$-$\alpha$ events from single-proton \big(\(^{20}\)Mg(\(\beta p\))\big) and single-alpha \big(\(^{20}\)Mg(\(\beta \alpha\))\big) events. These $p$-$\alpha$ events occupy a distinct region on a range vs. energy plot, located between the proton and alpha bands. Fig. \ref{fig:search_region} shows such a search region, extending 2 standard deviations in both range and energy.  By restricting our search to this region background events are reduced by more than 95\%. However, the remaining dataset is still substantial ($\sim$ 3.4 million events to sift through). Additional details on this method can be found in \cite{tyler_thesis}. 

\subsection{Data Preprocessing}\label{preprocessing}
The raw data within the search region undergoes several preprocessing steps. For each event, traces are recorded from the pads that were hit (see example in Fig. \ref{traces}). Each trace spans 512, 10 ns time bins, and the integral of the trace is proportional to the energy deposited on the pad. The peak's abscissa indicates the charge arrival time.

The trace data is stored in separate raw files, which are then merged and converted to HDF5 files for analysis \cite{10.1145/1966895.1966900}. During this conversion, channel signals are mapped to the pad plane layout. A peak detection algorithm identifies the peak time for each channel, and the detector's drift time is used to extract the z-coordinate of the charge deposition. The resulting HDF5 file contains, for each event, the x, y, z hit pattern, peak charge per pad, pad hit times, and the full trace data (512 elements).

These HDF5 files are then used for various analyses. In this case, outlier removal is applied using DBSCAN (Density-Based Spatial Clustering of Applications with Noise) to filter noise from random pad fires or beta particles \cite{10.1145/3068335}. Additionally, events are filtered based on integrated charge and track length thresholds, removing nonphysical events caused by detector noise, such as sparks. The x, y data is then used to generate 2D projections of particle tracks, with charge normalization—using the maximum pad charge per event—enabling clear Bragg peak detection, independent of the total event energy.

Additionally, for each event, the time projection is generated by summing all pad traces, and a Savitzky-Golay filter is applied to smooth the resulting 1D data. This filter fits adjacent data points with a low-degree polynomial to smooth the data \cite{5888646}.

\subsection{Custom Bounding Box Adjustments}\label{custom}
The bounding boxes undergo Non-Maximum Suppression (NMS) to filter out redundant boxes. However, false positives can still remain. Given that in this context Bragg peaks are being detected, there are additional physical constraints that can be considered. These constraints guide further refinements to the bounding boxes, ensuring that the detected Bragg peaks are not only distinct from one another but also physically reasonable.

First, Bragg peak boxes that are nested or overlapping to any degree are completely removed, retaining only the box with the highest confidence score. This step helps in reducing false positives by eliminating boxes that likely represent the same physical object but were detected with slightly different boundaries. It may be the case that Bragg peaks overlap for events very perpendicular relative to the pad plane, but in those instances the time projection will display a strong double peak, and thus will be flagged by the 1D peak finder. 

Another critical custom adjustment is the application of a separation threshold between Bragg peak boxes based on the size of the track bounding box. Bragg peaks that are detected too close together---closer than a threshold distance proportional to the size of the track---are considered unphysical and one of the boxes is removed. This adjustment ensures that the detected Bragg peaks are not only distinct but also realistically spaced according to the expected physical behavior of the event. By enforcing physical constraints and refining the bounding boxes, we improve the precision of our detections.

\subsection{2D Object Detection and 1D Peak Finding for Event Selection}
After 2D projections of all events in a given 2 standard deviation search region are generated for every experimental run (this corresponds to between 10,000-30,000 images for each region for several hundred runs), an object detector trained on parameter-varied simulated images is deployed on this data. (For information on the hardware used for the CNN deployment refer to \ref{hardware}). 

After prediction, and bounding box adjustments (see Sec. \ref{custom}) images with two or more Bragg boxes are flagged, and the images with the predicted bounding boxes overlaid are saved. These images can be used to scrutinize the performance of the the object detector.

The object detector can be deliberately set with a low confidence score threshold ($\sim$0.2) to maximize sensitivity and ensure that no potential two-particle events are overlooked, even those far outside the energy regime of primary interest. While this approach ensures that no two-particle events are missed, it also raises the number of false positives. However, the object detector can be deployed with a significantly higher threshold ($\sim$0.8) to identify only strong candidate p-alpha events near the energy-sharing region of interest, offering much greater precision.

In parallel with the deployment of the object detector, the smoothed 1D time projections for each event are fed into a the \texttt{scipy} peak-finding algorithm. The algorithm is configured with a threshold of 1000, a minimum distance of 20 ns between peaks, and a prominence value of 1 to detect significant peaks. Any event flagged by the peak-finding algorithm is cross-referenced against the list of events already flagged by the object detector, and duplicates are eliminated. The purpose of this algorithm is to identify events where the particle tracks are nearly perpendicular to the pad plane, resulting in a characteristic double peak in the time projection but not in the 2D track image. This ensures that rare two-particle events are captured that may be challenging or impossible to detect through the object detector alone.

\section{Results}\label{performance}
\subsection{Evaluation Metrics and Importance of Recall}
To evaluate the efficacy of the rare event detection framework it is compared to the performance of human evaluators. Performance for both is evaluated using the precision, recall, and F1 score metrics \cite{manning2009introduction}. It's important to note here that the most critical performance metric in this case is recall, as we are searching for extremely rare events and must prioritize minimizing missed detections, even at the cost of false positives.

\begin{figure*}
\centering
\includegraphics[width=0.8\textwidth]{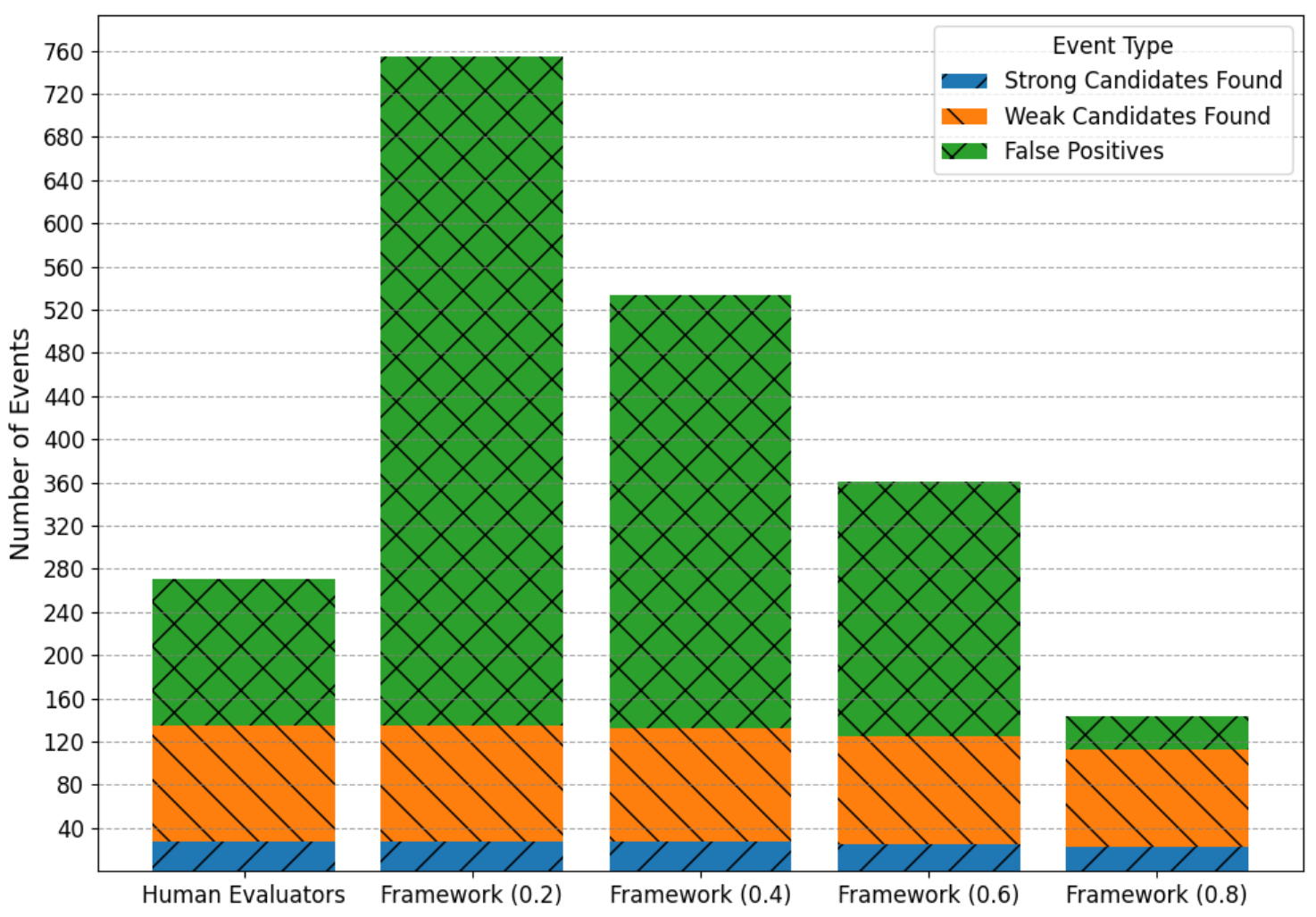}
\caption{Comparison of events found by human evaluators and the detection framework at various thresholds. The number in parentheses represents the threshold placed on the confidence score of the object detector. At the 0.2 threshold, the framework achieved 100\% recall on both strong and weak candidates. At the 0.4, the framework achieved 100\% recall on the strong candidates and identified 105 of the 107 weak candidates, resulting in a 98\% recall on weak events.}
\label{found}
\end{figure*}

\begin{figure*}
\centering 
\includegraphics[width=0.8\textwidth]{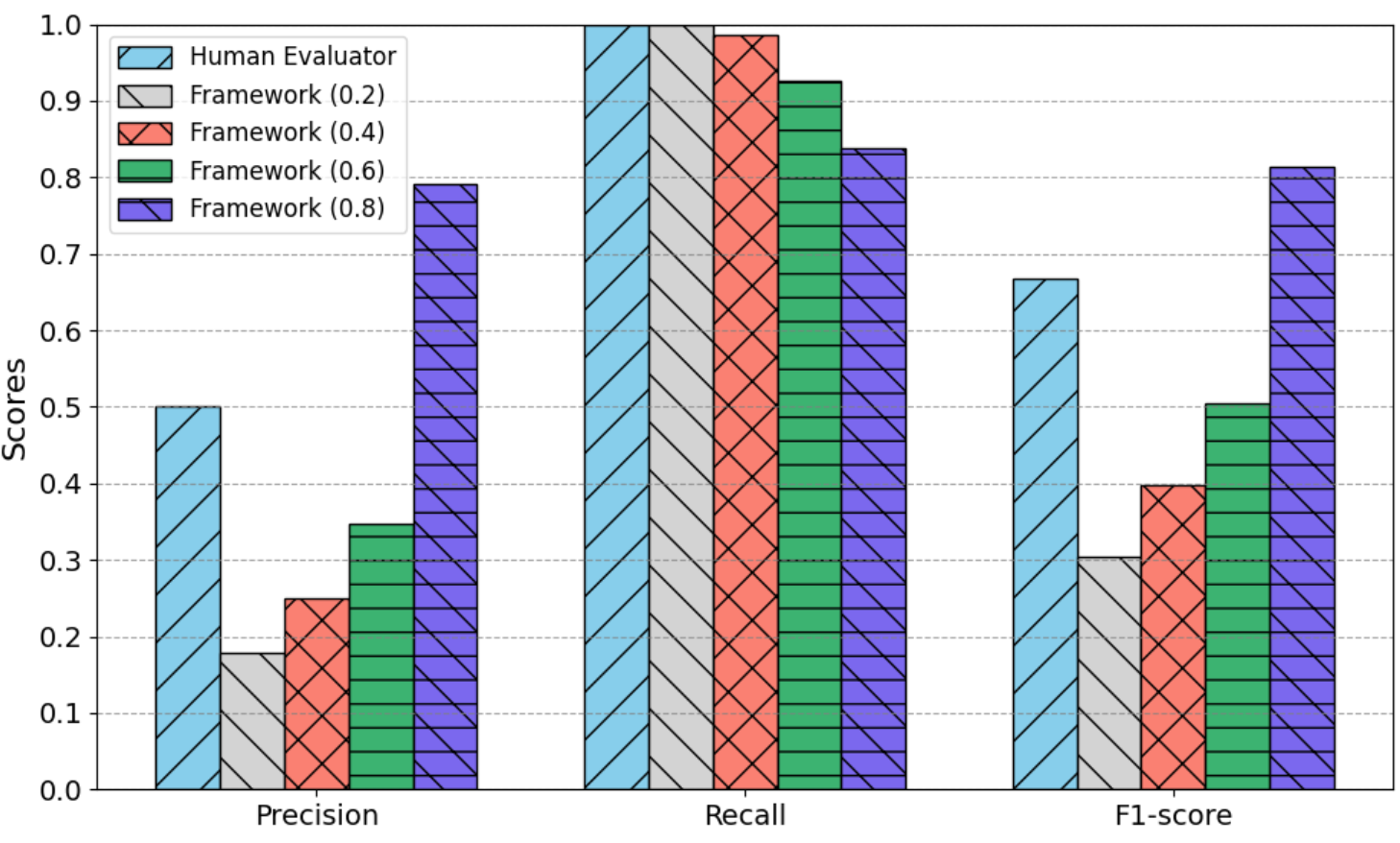} 
\caption{Comparison of performance metrics (Precision, Recall, F1-score) between human evaluators and the detection framework on all candidate events (strong and weak). The detection framework is evaluated using four different confidence score thresholds (0.2, 0.4, 0.6, 0.8). Human Evaluators: Precision = 0.498, Recall = 1.000, F1-score = 0.665.} 
\label{metrics} 
\end{figure*}

After an event is flagged by the evaluators, a separate analysis of the 3D data can indicate if a flagged event is a strong candidate, meaning it is likely to be a proton-alpha event near the energy-sharing region of interest, such as \textsuperscript{20}Mg($\beta p \alpha$) or \textsuperscript{21}Mg($\beta p \alpha$). Additionally, given the clear signature in a two-particle event, recall for the human evaluators is expected to be 100\%, but this manual examination requires such significant human effort that it is not feasible for an evaluator to go through more than a small fraction of the overall data. Thus, the goal of the detection framework is to filter out events with the features corresponding to a two-particle decay. Then subsequent analysis can be performed to confirm the isotopic origin of the two-particle event.

\subsection{Human Evaluation Methodology}

To establish a benchmark for the detection framework, human evaluators were tasked with manually inspecting a subset of the experimental data. Two evaluators, both members of the research group, independently reviewed composite images of $\sim$25,000 events. These composite images included both 2D track projections and 1D time projections for each event. Evaluators were instructed to flag any event that could even remotely be considered a two-particle event. This cautious approach aimed to ensure that no event of interest would be missed, accounting for the possibility of unknown two-particle decays with very different energy-sharing than the \textsuperscript{20}Mg($\beta p \alpha$) events of interest.

The flagged events from both evaluators were merged to create the \textit{initial flagged set}. In cases where one evaluator flagged an event but the other did not, the event was included in the flagged set to avoid omissions. However, discrepancies between the evaluators were minimal, and every discrepancy involved a false positive. 

Following the initial assessment by the evaluators, a more detailed analysis was conducted on the \textit{initial flagged set}, involving examinations of the raw traces and full 3D tracks with energy deposition for each flagged event. This analysis enabled the identification and removal of false positives. False positives can arise from various factors, the most common being pad noise, where a noisy pad mimics a small amount of particle energy deposition at one end of the track. Scattering events can also create the illusion of double Bragg peaks, as can the chance overlap of separate events, all of which further contribute to false positives.

The resulting dataset, referred to as the \textit{cleaned set}, was further divided into two groups: \textit{strong candidates} and \textit{weak candidates}. Strong candidates are likely to correspond to \textsuperscript{20}Mg($\beta p \alpha$) or \textsuperscript{21}Mg($\beta p \alpha$), while weak candidates may represent other two-particle events.

To classify events as strong or weak candidates, the energy of each event was projected onto its 3D trajectory using a kernel density estimator with Gaussian kernels. This approach can provide insights into the energy-sharing dynamics of multi-particle events. However, the method is sensitive to the choice of bandwidth and lacks statistically robust uncertainty quantification. As such, both strong and weak candidates remain classified as "candidates." Efforts to refine this classification are ongoing, including the exploration of the MCMC method for statistically quantifiable energy-sharing analysis \ref{phys_motivation}.

\begin{figure*}[h]
\centering
\includegraphics[width=0.9\textwidth]{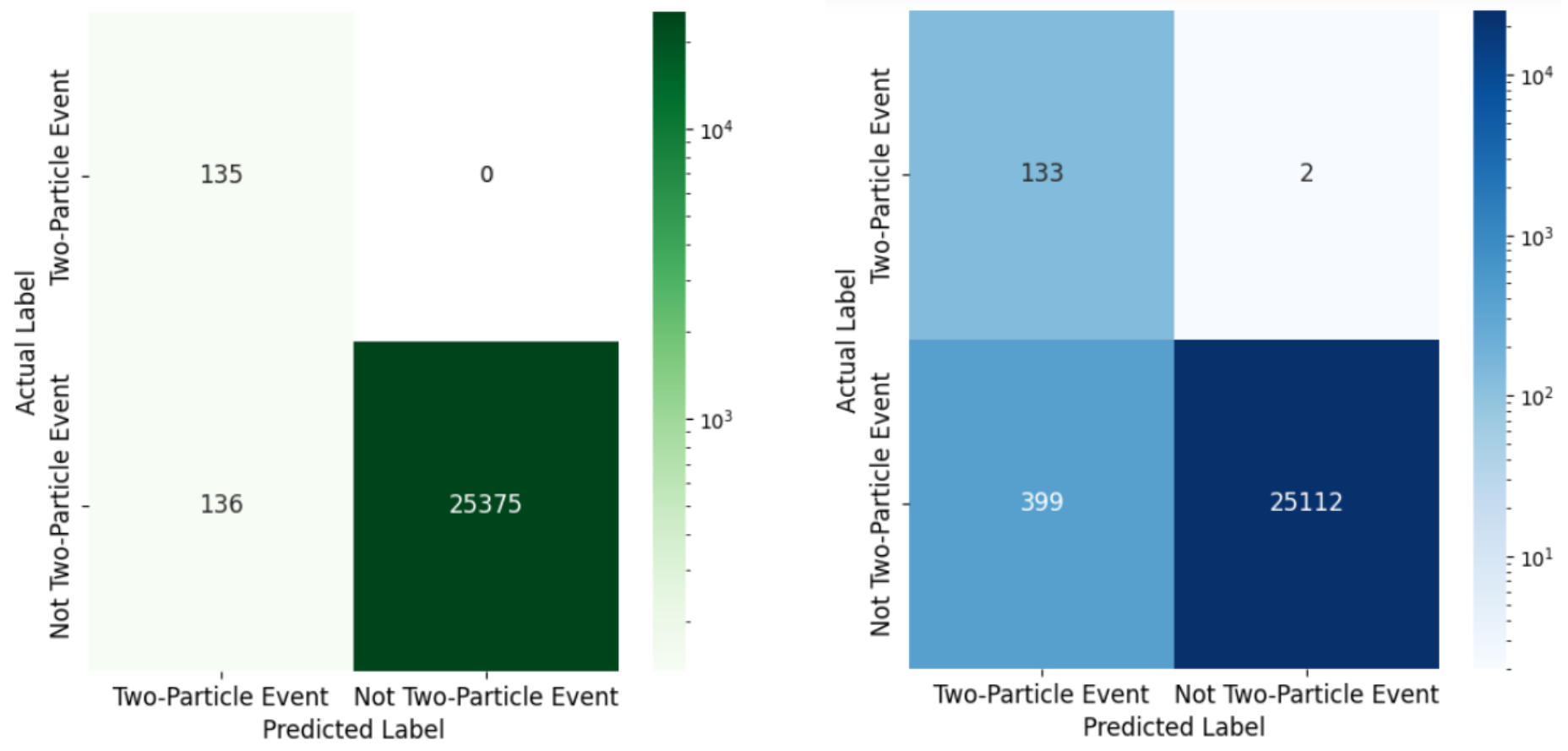}
\caption{Confusion matrices comparing the performance of human evaluators (left) and the detection framework (right) for all candidate events (strong and weak). The detection framework used a 0.4 classification score threshold for the object detector. Values along the diagonal indicate correctly classified instances, while off-diagonal values represent misclassifications (false positives in the bottom left, and missed events of interest in the top right).}
\label{confusion}
\end{figure*}

\subsection{Performance Comparison and Main Findings}

The human evaluators initially flagged 271 events as potential two-particle decays. The \textit{cleaned set} comprised 135 true positive events, which included 28 strong candidates and 107 weak candidates. The remaining 136 events were false positives. 

The detection framework was applied using varying object detector confidence thresholds of 0.2, 0.4, 0.6, and 0.8, each offering a distinct balance between recall and precision. At the 0.2 threshold, the framework achieved 100\% recall on both strong and weak candidates, successfully identifying all 28 strong candidates and all 107 weak candidates. Although this threshold resulted in a higher number of false positives, it is particularly noteworthy that the framework did not miss any candidate events. This performance is even more significant considering that the object detector was trained solely on parameter-varied simulations of \textsuperscript{20}Mg($\beta p \alpha$). The broad parameter space captured in these simulations appears to enable the detector to generalize effectively, identifying all potential two-particle events within this specific energy regime.

At the 0.4 threshold, the framework maintained a 100\% recall on the strong candidates and achieved 98\% recall on the weak candidates, while reducing the number of false positives by approximately 30\% compared to the 0.2 threshold. This highlights the framework’s adaptability—prioritizing both sensitivity and specificity without sacrificing key event detection.

At the higher threshold of 0.8, while recall decreased slightly, the framework demonstrated a highly balanced performance with precision, recall, and F1-scores around 0.8. Notably, this threshold produced 78\% fewer false positives than the human evaluators, suggesting that the framework can also offer a more precise and consistent detection process. 

The framework’s ability to achieve 100\% recall on all strong candidates across multiple thresholds speaks to its power in identifying rare events without omission—a crucial requirement in the given science case. The effectiveness of the framework is illustrated in Fig. \ref{found}, which compares the number of strong candidates, weak candidates, and false positives identified by both the human evaluators and the framework.

For completeness, the performance metrics for both the human evaluators and the detection framework for all candidate events (strong and weak) are summarized in Fig. \ref{metrics}. Additionally, confusion matrices for both methods are presented in Fig. \ref{confusion}, where the object detector used a threshold of 0.4 in the framework example.


\section{Conclusion and Outlook}

This study presents a rare event detection framework that significantly enhances the identification of two-particle decay events in a TPC. By training an object detector on as few as 200 parameter-varied simulated images of \textsuperscript{20}Mg($\beta p \alpha$) decays, combined with 1D peak detection on the time projections, the framework achieved 100\% recall on strong candidate events across multiple thresholds—successfully identifying all 28 critical events flagged by human evaluators. This is particularly noteworthy given the inherent challenges in detecting extremely rare events within vast datasets.

The framework can match human performance in recall while significantly surpassing it in efficiency by processing data at a scale impractical for manual analysis. At lower thresholds, higher false positives are observed, which is an acceptable trade-off in the context of rare event detection, where minimizing missed detections is paramount. At higher thresholds, such as 0.8, the framework achieves balanced performance, reducing false positives by up to 78\% compared to human evaluators. The effectiveness of this framework enables human experts to allocate their time more efficiently, focusing on detailed analysis of promising events rather than labor-intensive manual inspection of large datasets.

A key advantage of the framework is its minimal reliance on extensive training data. The success with a limited set of simulated images suggests that the simulations effectively capture the necessary parameter space to generalize well to experimental data. This reduces the burden of generating large labeled datasets, which is often a significant bottleneck in developing machine learning models for rare event detection.

Another significant benefit is the framework's potential for real-time deployment. With appropriate hardware optimizations (as detailed in \ref{hardware}), the detection system can be integrated directly into experimental setups to analyze data as it is collected. This real-time capability enables immediate feedback and adjustments during experiments, enhancing overall efficiency and resource utilization.

Although this framework was successfully applied to data from FRIB Experiment 21072, issues related to the dataset quality and experimental conditions limited its effectiveness for achieving the astrophysical constraints of interest \ref{phys_motivation}. The integrated \textsuperscript{20}Mg beam intensity from  experiment 21072 was a factor-of-17 lower than what was approved by PAC1, which hindered the ability to achieve the required statistical precision for astrophysical constraints on the \textsuperscript{15}O($\alpha$, $\gamma$)\textsuperscript{19}Ne reaction rate. Additionally, there was significant \textsuperscript{21}Mg contamination in the beam. While the \textsuperscript{21}Mg component was useful for commissioning the detection framework and methods to search for $\beta p \alpha$ events (as detailed in this paper), it introduced significant background to the \textsuperscript{20}Mg events of interest. 

Looking forward, the PAC3 approved FRIB Experiment 25058 will repeat this measurement with a higher-intensity, pure \textsuperscript{20}Mg beam, allowing for a finite measurement of the \textsuperscript{20}Mg branching ratio. The framework outlined in this study will play a critical role in identifying events of interest in the upcoming dataset, with MCMC analyses used to confirm the isotopic origin of detected events \ref{phys_motivation}. By achieving high recall with minimal training data and enabling real-time processing, this approach promises to enhance the efficiency and effectiveness of this and other future experimental campaigns. Additionally, the methodologies developed here have broader implications and could be adapted to not only other TPCs, but other image-generating detectors requiring rare event or rare feature detection in other fields such as astrophysics and particle physics.

\section{Acknowledgments}
This work was supported by the U.S. National Science Foundation under Grants No. PHY-1102511, PHY-1565546, PHY-1913554, PHY-1811855, PHY-2209429, PHY-2310059, PHY-1848177 (CAREER) (Mississippi State), and CCF-2212065, the  Institute for Basic Science (IBS) of the Republic of Korea under Grant No. IBS-R031-D1, the International Technology Centre Pacific (ITC-PAC) under Contract No. FA520919PA138, and the U.S. Department of Energy, Office of Science, under award No. DE-SC0016052, DE-SC0023529, and DE-SC0024587, and Office of Nuclear Physics under Contract No. DE-AC05-00OR22725 (ORNL). 

\appendix

\section{Hardware}
\label{hardware}

For efficiency in CNN deployment, particularly for the computationally intensive tasks of object detection and image analysis, a powerful GPU node is utilized. This machine is equipped with the following hardware specifications:

\begin{itemize} \item \textbf{CPU:} Dual AMD EPYC 7763 processors with 128 physical cores (256 threads total), optimized for high-performance parallel computing. \item \textbf{Memory:} Includes 512 MiB of L3 cache for efficient data processing during memory-intensive tasks. \item \textbf{GPU:} Four NVIDIA RTX A6000 GPUs, each with 48 GB GDDR6 memory, ideal for AI training and deep learning. \item \textbf{CUDA:} Runs CUDA Version 12.2 to accelerate CNN model training and inference. \item \textbf{OS/Software:} 64-bit Linux OS with NVIDIA driver 535.183.01, optimized for deep learning workflows. \end{itemize}

This hardware configuration allows for the efficient execution of deep learning models, ensuring that the computational demands of training and inference are met with speed and reliability. The combination of high core count CPUs and powerful GPUs provides the necessary parallel processing power to handle the large-scale image analysis tasks required in this project.

 \bibliographystyle{elsarticle-num} 
 \bibliography{r}





\end{document}